# Universities through the Eyes of Bibliographic Databases: A Retroactive Growth Comparison of Google Scholar, Scopus and Web of Science[1]

**Enrique Orduna-Malea**
enorma@upv.es
*Universitat Politècnica de València*

**Selenay Aytac**
Selenay.Aytac@liu.edu
*Long Island University*

**Clara Y. Tran**
yuet.tran@stonybrook.edu
*Stony Brook University*

**Abstract**
The purpose of this study is to ascertain the suitability of GS's url-based method as a valid approximation of universities' academic output measures, taking into account three aspects (retroactive growth, correlation, and coverage). To do this, a set of 100 Turkish universities were selected as a case study. The productivity in Web of Science (WoS), Scopus and GS (2000 to 2013) were captured in two different measurement iterations (2014 and 2018). In addition, a total of 18,174 documents published by a subset of 14 research-focused universities were retrieved from WoS, verifying their presence in GS within the official university web domain. Findings suggest that the retroactive growth in GS is unpredictable and dependent on each university, making this parameter hard to evaluate at the institutional level. Otherwise, the correlation of productivity between GS (url-based method) and WoS and Scopus (selected sources) is moderately positive, even though it varies depending on the university, the year of publication, and the year of measurement. Finally, only 16% out of 18,174 articles analyzed were indexed in the official university website, although up to 84% were indexed in other GS sources. This work proves that the url-based method to calculate institutional productivity in GS is not a good proxy for the total number of publications indexed in *WoS* and *Scopus*, at least in the national context analyzed. However, the main reason is not directly related to the operation of GS, but with a lack of universities' commitment to open access.

---

# 1. Introduction

Over the past few decades, using bibliometric data for the evaluation of universities performance became common practice (Mingers and Meyer 2017). The research strength of a particular university is usually measured by means of indicators related to research productivity, such as the number of published contributions that have been co-authored at least by one member affiliated to the university. According to this assumption, the 'quantity of publications' is directly related to the universities' commitment to research activities, leading to the production of scholarly outcomes (Ramsden 1994).

Both this metric 'quantity of publications' and its linked indicator 'research productivity' are extensively used in institutional research evaluation exercises and bibliometric studies. They are also included in the methodology of most domestic and international university rankings, although with different denominations; for example: Publications (*Leiden Ranking*, *US News' Best Global Universities*), Research Output (*Academic Ranking of World Universities*) or Research Productivity (*Times Higher Education World University Rankings*). Additionally, this indicator is indirectly used to measure other university ranking parameters such as citation-based research impact. Given the media influence of these rankings, it is not surprising that this indicator has been regularly monitored by higher education administrators (Amara, Landry and Halilem 2015).

However, the research productivity constitutes an indicator with limitations as a proxy of the universities' research strength. It depends on the quality and veracity of the affiliation data included either directly in the documents or indirectly in the metadata provided by the bibliographic databases. This excludes research activities and intellectual influences beyond the formal publication in scholarly journals.

Otherwise, research productivity depends on the coverage of the selected bibliographic databases and the quantity of published scholarly outputs. To date, the most widely used databases used to determine universities' research productivity are *Clarivate Analytics*' *Web of Science* (WoS) and *Elsevier*'s *Scopus* (along with their star premium products, *Incites* and *Scival*, respectively). Furthermore, *Alphabet*'s *Google Scholar (GS)*, *Microsoft Academic (MA)* and − more recently − *Digital Science*'s *Dimensions*, haven been incorporated as supplementary bibliographic databases available in the scientific information market.

Both *WoS* and *Scopus* are bibliographic databases that are equipped with the necessary functionalities to be used in bibliometric analysis (including institutional productivity). Their operating is based on Bradford's law of scattering (elitist selection of sources) and a semi-controlled management of the bibliographic references. However, this does not make them immune to errors (Franceschini, Maisano, and Mastrogiacomo 2016a; 2016b) while introducing a series of notable biases towards certain disciplines such as Natural Sciences, Medicine Physics, and related, and serious limitations towards sources written in languages other than English (Mongeon and Paul-Hus 2016), which may lead to less accurate representation of the research capacity of certain universities.

Contrariwise, GS is a dynamic, open, and uncontrolled database which covers all languages, typologies, and disciplines (Martín-Martín et al. 2016). However, some limitations of GS must be recognized as some technical requirements (Orduna-Malea et al. 2016) with data quality as well as search limitations to be used accurately as a Bibliometrics tool (Aguillo 2012; Delgado López-Cózar, Orduna-Malea, and Martín-Martín 2019). Otherwise, MA exhibits a higher coverage than *WoS* and *Scopus* while following a non-elitist approach, but lower than GS, and with a considerable number of publications with missing or wrong affiliation data (Ranjbar-Sahraei and Van Eck 2018). Finally, *Dimensions* currently shows a total coverage higher than *Scopus* (99,523,454 and 74,353,833 documents, respectively). However, institutional data seems to be lower (for example, *Scopus* offers 242,891 indexed documents for the *Massachusetts Institute of Technology* whereas *Dimensions* only offers 166,724), probably due to its dependence on *Crossref* metadata (Hook, Porter, and Herzog 2018; Orduna-malea and Delgado López-Cózar 2018).

The comparison of an institution's scholarly productivity in different bibliographic databases such as *WoS*, *Scopus,* and *GS* constitutes a required exercise to verify to the extent that the coverage and accuracy of these information products influence the results obtained. While strong correlations will reinforce the productivity of institutions, weak correlations may exhibit coverage bias. This comparison is however compromised, due to the general operating and specific features offered by the bibliographic databases.

For example, it is not possible to accurately retrieve the number of documents published by one university in *GS*. Instead of this, users can perform specific queries to retrieve the number of documents stored in the official university website, which represents a rough proxy. This url-based method has been long utilized by the *Ranking Web of Universities* (Aguillo, Ortega & Fernandez 2008), in which the number of total documents (or in pdf, doc, and ppt) deposited within the official university website and indexed by *GS.*

Beyond the accuracy and coverage of *GS*, the url-based proxy adds a new variable to the equation related to the dependence on institutional web policies. More specifically, the number of documents retrieved heavily depends on the role of institutional repositories in providing access to institutions scholarly output. That is, if documents are not deposited (or are deposited incorrectly, according to the *GS* rules), *GS* will not index them; being this issue is one of the main consequences of a dynamic and self-controlled academic search engine like *GS.*

Since documents can be indexed by means of other sources (personal websites, journal websites, institutional repositories of co-authors, academic social networking sites, etc.), the effect of wrong web policies may provoke institutional web invisibility (Arlitsch and O'Brian 2012; Orduna-Malea and Delgado López-Cózar 2015). This is a significant problem given the massive use of *GS* as a starting point for bibliographic research.

We can hypothesize that institutional productivity between bibliographic databases will correlate which means that the most productive universities are regardless of the source we use to measure them. In addition, we can also assume that the volume of data will be higher in *GS* because of its greater coverage. Then, any deviation from these two premises (strong correlation and greater data volume in *GS*) would indicate an underestimation of the productivity in the search engine. Consequently, *GS* would not be appropriate to measure universities' academic output but would be useful to detect institutional web invisibility.

It would be necessary however to carry out the analyses in time series, in order to determine whether anomalous values (deviated from the starting hypothesis) are due to one-time or systemic errors. In this sense, the dynamism of *GS* introduces another drawback, since the access of its crawlers to previously closed sources, the effectiveness of its parsers, as well as the digitalization of printed articles can generate a significant retroactive growth (De Winter, Zadpoor, and Dodou 2014), making the correlations dependent not only on the year of publication but also on the year of measurement.

Notwithstanding, the retroactive growth in *GS* has been scarcely treated in the literature (See Research Background). As a consequence, it is currently not possible to ascertain the suitability of *GS*'s url-based method as a valid proxy of universities' academic output measures, which constitutes the main objective of this work. To investigate this, the following research questions are addressed:

**RQ1.** What is the retroactive growth in *GS* compared to the equivalent of *WoS* and *Scopus*?

**RQ2.** Is there a correlation between the number of documents deposited on the university website and indexed by *GS*, and the number of articles published by these universities and indexed in *WoS* and *Scopus*? Does this correlation vary over time?

**RQ3.** What is the coverage of documents (institutionally published by a university and indexed in *WoS*) deposited in the university websites and indexed by *GS*?

## 2. Research background

*Google Scholar* is an academic search engine launched in November 18[th] 2004 by Mountain View's company *Google Inc.* (now *Alphabet*). The release of this product represented the work by Anurag Acharya and Alex Verstak, who realized that academic queries by researchers and students had specific patterns that differ from general queries so that they could automatically filter them to offer more appropriate results embedded in the general search engine.

The potential of this idea was however beyond what was initially planned and *GS* was developed as a separate search engine, whose public presentation almost concurred with the milestone release of another bibliographic database (*Scopus*), launched by *Elsevier* just few weeks before (Orduna-Malea et al. 2016). Whereas *Scopus* emerged with a clear intention of breaking the market monopoly of supervised and elitist bibliographic databases represented by the binomial *Web of Science*/*Journal Citation Reports*, *GS* came to represent different objectives and approaches from a complementary and emerging threat market, the academic search engines (Ortega 2014).

*GS* consisted of a specialized search engine oriented to the index and retrieval of academic contents available online worldwide, similar to other search products such as *MA* or *Semantic Scholar*, with the particularity that it incorporates *Google* general search engine technology.

The coverage of *GS* experienced an astonishing growth from its inception (Ortega 2014). Its coverage was estimated to contain over 170 million documents up to 2013 (Orduna-Malea et al. 2015), figure that has amounted to near the 400 million total records (Gusenbauer 2019; Delgado López-Cózar, Orduna-Malea, and Martín-Martín 2019). However, significant errors in the bibliographic data (records without year of publication, duplicated records) and limitations (lack of authority control) lead researchers to rely on dubious hit count estimates (Jacsó 2010; Orduna-Malea, Martín-Martín, and Delgado López-Cózar 2017).

In the presence of such constrains, the analysis of universities' academic productivity according to *GS* has been performed cautiously. Moskovkin, Delux and Moskovkina (2012) analyzed leading Czech and Germany universities whereas Orduna-Malea, Serrano-Cobos, and Lloret-Romero (2009) analyzed the Spanish university system. The latter authors tested the correlation between the research output from *Scopus* and the hit count estimates from *GS*. The results revealed that, despite finding some interrelationship between *Scopus* and *GS* concerning the productivity of institutions, there were large differences in the total results that override the latter as a valid reflection of university production. Notwithstanding, those results were obtained from data gathered back in 2009 and only from Spanish public universities. The evolution of institutional repositories, the subsequent improvement of *GS* database, and the consideration of other academic systems (likely to make different use of technology media) might change the state of affairs.

Similarly, Moskovkin (2009) finds absence of correlation between the indices of *WoS* and *GS* publications in his analysis of nine highly productive universities, linking these results to the bad web-presentation of publications for universities for which the ratio of GS/WoS is low.

The role of the retroactive growth in these correlations has not been treated so far. De Winter, Zadpoor and Dodou (2014) compare the retroactive and actual growth of *GS* versus *WoS* through a longitudinal analysis. However, this study is performed from the point of view of citation counts, not productivity. In any case, authors find that *GS* demonstrated a striking retroactive growth. Similarly, Harzing (2013; 2014) studied the growth of *GS* at the author-level and citation-side.

With the later release of the official *GS*'s derivative product *Google Scholar Citations* (November 2011), the bibliometric use of *GS* data expanded directly to the author-level adding a supplementary quality control on the user-side. *GS* released in 2015 an automatic institutional affiliation tool incorporated within each public academic profile, which aims to gather all authors belonging to one institution (*Google Scholar Citation Institutional Profiles*) with their corresponding citation counts.

This tool is currently used to rank universities according to the number of citations (http://webometrics.info/en/transparent). However, it suffers from certain inconsistencies that jeopardizes its use (Orduna-Malea et al. 2017). Mingers, O'Hanley and Okunola (2017) also evaluated the possibility of using *Google Scholar Citation Institutional Profiles* data to evaluate the university research of 130 UK institutions, obtaining credible rankings.

However, an analysis of the academic productivity of universities according to the current coverage of *GS*, and the effects of its retroactive growth in its correlation with other bibliographic databases such as *WoS* and *Scopus* has not been performed to date.

## 3. Method

The analysis is bounded to a national university system. In this sense, Turkey provides an appropriate testing bed. This educational system has experienced important changes in the last decade that may be reflected in the research output shown by the bibliographic databases.

The top 100 Turkish universities according to the *Ranking Web of Universities* were selected for the study (July 2014 edition). This ranking classifies universities according to their web performance (http://www.webometrics.info/en/Europe/Turkey). The official URL for each university website was gathered directly from this source.

The total amount of academic productivity data was obtained from *Scopus* and *WoS* (*Core Collection*). As many of the institutions in the sample were created recently (within the last 10 years), results were restricted to the period 2000-2013. As regards *GS*, the hit count estimates (number of documents stored within each official university website) were retrieved with the "site" search command applied to each official university web domain (e.g., site:istanbul.edu.tr), annually filtering the year of publication from 2000 to 2013 (a total of 14 queries). The three different search strategies are the following (Table 1):

**Table 1. Search strategies used in *Scopus*, *Web of Science* and *Google Scholar***

| Database | Search strategy |
|---|---|
| *Scopus* | AFFIL(University-Name) AND PUBYEAR = 2000 |
| *Web of Science* | OG=(University Name) AND PY= 2000 |
| *Google Scholar* | https://scholar.google.com/scholar?q=site:university-website-domain&hl=en&as_ylo=2000&as_yhi=2000 |

In order to respond RQ1 and RQ2, all data were manually retrieved in two separate iterations. The first sample was retrieved in December 2014, and second in March 2018. The retroactive growth rate (which presents a margin of four years) was calculated for each database as follows:

[1] $RAgr_{u,y} = \frac{P_{u,y,i2}}{P_{u,y,i1}} \times 100;$

Retroactive Growth rate of university "u" in the year of publication "y", where the first iteration "i1" is 2014 and the second iteration "i2" is 2018.

[2] $RAgr_y = \frac{\sum_{n=1}^{100} RAR_{u_n,y}}{n};$

Retroactive Growth rate of year "y" considering the average of the Retroactive Growth Rate from a set of 100 Turkish universities.

[3] $RAgr_t = \frac{\sum_{n=1}^{14} RAR_{y_n}}{n};$

Total Retroactive Growth rate considering the Retroactive Growth Rate of a set of years of publication (from 2000 to 2013).

After this, the productivity results obtained from *Scopus*, *WoS* and *GS* were correlated to each other in each of the two iterations (2014 and 2018) in order to find data similarity. Since web data presents a skewed distribution, *Spearman* (α=0.05) was applied in all calculations.

To answer the RQ3, a sample of the 14 universities was taken: the top ten institutions with the highest scientific output in *WoS* in 2013 (See Table 7), and four additional universities (*Boğaziçi Üniversitesi*, *Erciyes Üniversitesi*, *Gebze Teknik Üniversitesi*, and *İzmir Yüksek Teknoloji Üniversitesi*). The latter institutions were selected as they have been considered as research-focused institutions (Aytac 2010).

A total of 18,174 documents published in 2013 (measured in 2018) and co-authored at least by one author affiliated to one of the 14 universities were retrieved from *WoS*. For each document, a query in *GS* was performed (enclosing the title in quotation marks and using 'allintitle' advance query when necessary) in order to check whether it was indexed in the database.

If the article was located, then it was checked if the record was complete or 'citation' type (a record not directly indexed in *GS* but appearing in the reference set of a document already indexed). Finally, the presence of the university website as a source of the record was verified, both as a primary version (appearing directly in the search engine results page) or supplementary version (appearing through 'all x versions' button below each record).

Some titles could not be recovered because they had too generic titles or simply the corresponding field of the bibliographic record exported from *WoS* was empty (or just named 'Untitled'). Those documents were discarded from the analysis.

The search of documents in *GS* through the Title field yielded three additional shortcomings (Figure 1):

a) The presence of alternative titles (documents originally written in Turkish but directly translated in the title field provided by *WoS*). In some occasions, the document was only discovered using the original title.
b) Errors were in the transcription of the title both in *WoS* and *GS*.
c) Document versions were not properly linked in *GS*. In those cases when the same document was located in different *GS* records, each one was treated independently when assigning them as Primary or Supplementary version.

**Figure 1. Overview of the general limitations of Title Search in *Google Scholar***

A manual verification of each document (checking indexation and *GS* version condition) was therefore needed. This task was distributed in an equitable manner by three co-authors. To carry out this process, an inter-coder reliability test was performed in order to assure the accuracy of results. Each author (coder) repeated the procedure for the same set of data and the results were discussed among the three authors to assess the level of inter-coder reliability. The procedure repeated multiple times including the beginning, middle, and towards the end of the data collection period to obtain very high inter-coder reliability.

## 4. Results

**RQ 1. Retroactive growth in *Google Scholar*, *Scopus* and *Web of Science***

The total number of documents (without eliminating duplicates, i.e., documents co-authored by two or more universities within the sample) published by the 100 Turkish universities from 2000 to 2013 amounts to 344,615 when measured in 2014, and to 353,879 when measured in 2018, which corresponds to a total retroactive growth rate of 2.7%. In the case of *Scopus* this growth rate is higher (4.1%) and, surprisingly, achieves a negative value for *GS* (-1.2%).

We can observe the evolution of the retroactive growth if we disaggregate productivity yearly (Table 2 and 3). *WoS* exhibits a minor negative growth at the beginning of the period (probably due to retractions and bibliographic errors), and positive growth in the last years (probably due to the intake of new documents not indexed yet during the first iteration). *Scopus* shows a similar pattern in the last years although it clearly differs in the first years (positive growth probably due to an increase of the coverage of indexed sources). Finally, *GS* presents an unpredictable pattern (strong positive and negative growths) over the time (due to its dynamism dependent on the information available online and the accuracy of its parsers and crawlers).

**Table 2. Retroactive growth rate in *Google Scholar*, *Scopus* and *Web of Science* broken down by year of publication**

| YEAR | WEB OF SCIENCE | | | | SCOPUS | | | | GOOGLE SCHOLAR | | | |
|---|---|---|---|---|---|---|---|---|---|---|---|---|
| | 2014 | 2018 | DIF. | % | 2014 | 2018 | DIF. | % | 2014 | 2018 | DIF. | % |
| 2000 | 7516 | 7476 | -40 | -0.5 | 7931 | 8311 | 380 | 4.8 | 1507 | 1788 | 281 | 18.6 |
| 2001 | 9124 | 9093 | -31 | -0.3 | 9209 | 9998 | 789 | 8.6 | 2227 | 2217 | -10 | -0.4 |
| 2002 | 11942 | 11890 | -52 | -0.4 | 12243 | 12845 | 602 | 4.9 | 3331 | 2933 | -398 | -11.9 |

| YEAR | WEB OF SCIENCE | | | | SCOPUS | | | | GOOGLE SCHOLAR | | | |
|---|---|---|---|---|---|---|---|---|---|---|---|---|
| 2003 | 14924 | 14885 | -39 | -0.3 | 16571 | 17167 | 596 | 3.6 | 3942 | 3841 | -101 | -2.6 |
| 2004 | 18586 | 18498 | -88 | -0.5 | 20608 | 20753 | 145 | 0.7 | 5338 | 5141 | -197 | -3.7 |
| 2005 | 20334 | 20248 | -86 | -0.4 | 23263 | 23317 | 54 | 0.2 | 6544 | 5328 | -1216 | -18.6 |
| 2006 | 23182 | 23318 | 136 | 0.6 | 25869 | 26210 | 341 | 1.3 | 6783 | 7029 | 246 | 3.6 |
| 2007 | 27816 | 28331 | 515 | 1.9 | 28798 | 29234 | 436 | 1.5 | 8444 | 8585 | 141 | 1.7 |
| 2008 | 28858 | 29278 | 420 | 1.5 | 29304 | 30006 | 702 | 2.4 | 7597 | 8498 | 901 | 11.9 |
| 2009 | 32934 | 33705 | 771 | 2.3 | 33540 | 34229 | 689 | 2.1 | 8338 | 9282 | 944 | 11.3 |
| 2010 | 34072 | 35360 | 1288 | 3.8 | 35490 | 36753 | 1263 | 3.6 | 9996 | 9997 | 1 | 0.0 |
| 2011 | 35766 | 37324 | 1558 | 4.4 | 37542 | 38969 | 1427 | 3.8 | 11930 | 10751 | -1179 | -9.9 |
| 2012 | 38523 | 40548 | 2025 | 5.3 | 38008 | 41461 | 3453 | 9.1 | 11996 | 10407 | -1589 | -13.2 |
| 2013 | 41038 | 43925 | 2887 | 7.0 | 40286 | 44260 | 3974 | 9.9 | 11409 | 12434 | 1025 | 9.0 |
| **TOTAL** | **344615** | **353879** | **9264** | **2.7** | **358662** | **373513** | **14851** | **4.1** | **99382** | **98231** | **-1151** | **-1.2** |

Table 3 provides complementary statistical data related to the distribution of the difference in productivity between the two iterations (raw retroactive growth rate) for the 100 universities, specifically the Mean of the distribution, the percentage (percentage growth rate), the maximum and minimum difference obtained in one particular observation, and the Statistical Deviation (SD), by year. As we can see, the dispersion of data is significant, not only between years, but also within the same year (wide statistical range of the difference between universities).

**Table 3. Statistical data related to the distribution of the raw retroactive growth rate in *Google Scholar*, *Scopus* and *Web of Science* broken down by year of publication**

| YEAR | WEB OF SCIENCE | | | | SCOPUS | | | | GOOGLE SCHOLAR | | | |
|---|---|---|---|---|---|---|---|---|---|---|---|---|
| | MEAN | MAX | MIN | SD | MEAN | MAX | MIN | SD | MEAN | MAX | MIN | SD |
| 2000 | -0.4 | 23 | -27 | 4 | 3.8 | 87 | -14 | 11.3 | 2.8 | 170 | -373 | 49.0 |
| 2001 | -0.3 | 23 | -19 | 3.7 | 7.9 | 93 | -6 | 15.4 | -0.1 | 355 | -422 | 68.4 |
| 2002 | -0.5 | 23 | -28 | 4.6 | 6 | 130 | -191 | 26.5 | -4 | 409 | -693 | 91.5 |
| 2003 | -0.4 | 27 | -27 | 5.1 | 6 | 136 | -73 | 20.1 | -1 | 452 | -548 | 83.0 |
| 2004 | -0.9 | 21 | -35 | 5.3 | 1.5 | 134 | -90 | 20.2 | -2 | 629 | -683 | 110.2 |
| 2005 | -0.9 | 16 | -30 | 5.1 | 0.5 | 124 | -97 | 21.9 | -12.2 | 474 | -842 | 115.4 |
| 2006 | 1.4 | 31 | -112 | 13.4 | 3.4 | 131 | -97 | 23.8 | 2.5 | 984 | -677 | 161.4 |
| 2007 | 5.2 | 61 | -32 | 11.3 | 4.4 | 145 | -73 | 22.9 | 1.4 | 901 | -739 | 147.4 |
| 2008 | 4.2 | 58 | -22 | 9.9 | 7 | 157 | -88 | 24.4 | 9 | 903 | -665 | 158.5 |
| 2009 | 7.7 | 70 | -18 | 13.4 | 6.9 | 160 | -99 | 27.5 | 9.4 | 942 | -586 | 175.7 |
| 2010 | 12.9 | 112 | -11 | 19.7 | 12.6 | 149 | -74 | 24.7 | 0 | 1091 | -694 | 202.0 |
| 2011 | 15.6 | 97 | -10 | 18.2 | 14.3 | 213 | -508 | 65.8 | -11.8 | 955 | -1290 | 210.2 |
| 2012 | 20.3 | 146 | -6 | 22.1 | 34.5 | 369 | -84 | 56.2 | -15.9 | 707 | -1215 | 175.6 |
| 2013 | 28.9 | 197 | -15 | 35.2 | 39.7 | 409 | -76 | 59.1 | 10.3 | 881 | -610 | 157.0 |

If we pay our attention to the 10 most productive universities according to *WoS* (published in 2013 and measured in 2014), we can observe how their behavior patterns differ (Table 4). For example, *Istanbul University* exhibits a huge decrease (a loss of 10,037 documents) while *Istanbul Technical University* an outstanding increase (9,068 documents more). What is more, some universities present different behaviors according to the database. *Gülhane* disappears in *GS* (0 documents indexed in the second iteration) although in *Scopus* shows an increase of 1889.

**Table 4. Retroactive growth rate in *Google Scholar*, *Scopus* and *Web of Science* broken down by University**

| YEAR | WEB OF SCIENCE | | | | SCOPUS | | | | GOOGLE SCHOLAR | | | |
|---|---|---|---|---|---|---|---|---|---|---|---|---|
| | 2014 | 2018 | DIF | % | 2014 | 2018 | DIF | % | 2014 | 2018 | DIF | % |

| | | | | | | | | | | | |
|---|---|---|---|---|---|---|---|---|---|---|---|
| Istanbul | 20944 | 20982 | 38 | 0.2 | 21986 | 21071 | -915 | -4.2 | 10322 | 285 | -10037 | -97.2 |
| Hacettepe | 20502 | 20907 | 405 | 2.0 | 20062 | 20381 | 319 | 1.6 | 3082 | 2649 | -433 | -14.0 |
| Gazi | 13933 | 14170 | 237 | 1.7 | 14990 | 15376 | 386 | 2.6 | 6050 | 4833 | -1217 | -20.1 |
| Ege | 13825 | 14091 | 266 | 1.9 | 14141 | 14246 | 105 | 0.7 | 1515 | 226 | -1289 | -85.1 |
| Ankara | 16022 | 16312 | 290 | 1.8 | 16674 | 16840 | 166 | 1.0 | 4830 | 4094 | -736 | -15.2 |
| Middle East Tech. | 13094 | 13931 | 727 | 5.5 | 14617 | 16694 | 2077 | 14.2 | 7484 | 5833 | -1651 | -22.1 |
| Istanbul Tech. | 11385 | 11876 | 491 | 4.3 | 13250 | 13873 | 623 | 4.7 | 2009 | 11077 | 9068 | 451.4 |
| Gülhane | 8011 | 7980 | -31 | -0.4 | 6568 | 8457 | 1889 | 28.8 | 31 | 0 | -31 | -100.0 |
| Erciyes. | 7711 | 7860 | 149 | 1.9 | 8330 | 8595 | 265 | 3.2 | 1382 | 624 | -758 | -54.8 |
| Dokuz Eylül | 8926 | 9142 | 216 | 2.4 | 9040 | 9303 | 263 | 2.9 | 1822 | 6255 | 4433 | 243.3 |

With the aim of testing the influence of *Istanbul University* in the overall results, we replicated the retroactive growth rate without this institution. The results show a slight increase of this value both in *WoS* (from +2.7 to +2.9) and *Scopus* (from +4.1 to +4.7), and a very big increase in *GS* (from -1.2 to +10), a value more in line with the initially expected.

The average growth rate per year (from 2000 to 2013) is also offered (Table 5), supplemented by some statistics (Mean, Maximum difference, Minimum Difference, Standard Deviation).

**Table 5. Statistical data related to the distribution of the raw retroactive growth rate in *Google Scholar*, *Scopus* and *Web of Science* broken down by University**

| UNIVERSITY | WOS | | | | SCOPUS | | | | GOOGLE SCHOLAR | | | |
|---|---|---|---|---|---|---|---|---|---|---|---|---|
| | MEAN | MAX | MIN | SD | MEAN | MAX | MIN | SD | MEAN | MAX | MIN | SD |
| Istanbul | 2.7 | 139 | -112 | 54.5 | -65.4 | 33 | -101 | 44.2 | -716.9 | -373 | -1290 | 257.4 |
| Hacettepe | 28.9 | 112 | -4 | 39.0 | 22.8 | 73 | -10 | 25.3 | -30.9 | 527 | -178 | 171.6 |
| Gazi | 16.9 | 94 | -14 | 32.3 | 27.6 | 143 | -6 | 38.2 | -86.9 | 37 | -465 | 129.0 |
| Ege | 19.0 | 60 | 0 | 21.7 | 7.5 | 83 | -35 | 32.5 | -92.1 | -6 | -298 | 83.0 |
| Ankara | 20.7 | 91 | -5 | 29.9 | 11.9 | 52 | -18 | 22.4 | -52.6 | 50 | -176 | 59.5 |
| Middle East Tech. | 51.9 | 166 | -5 | 58.6 | 148.4 | 255 | 87 | 44.7 | -117.9 | 30 | -295 | 90.3 |
| Istanbul Tech. | 35.1 | 130 | -13 | 45.4 | 44.5 | 140 | 18 | 31.7 | 647.7 | 1091 | 147 | 280.0 |
| Gülhane | -2.2 | 20 | -18 | 10.7 | 134.9 | 409 | 23 | 120.4 | -2.2 | 0 | -19 | 5.0 |
| Erciyes. | 10.6 | 45 | -1 | 13.6 | 18.9 | 122 | -29 | 45.4 | -54.1 | -1 | -98 | 34.7 |
| Dokuz Eylül | 15.4 | 77 | -3 | 22.2 | 18.8 | 119 | -7 | 40.7 | 316.6 | 755 | 54 | 239.7 |

The strength of the variations is in general smaller in *WoS*, medium in *Scopus*, and large in *GS*, were we detect extreme values. For example, *Istanbul University* suffered a maximum drop of published documents in 2011, when the institution shows 1290 fewer documents when measuring in 2018 with respect to the same measurement in 2014. On the contrary, *Istanbul Technical University* increases 1091 documents for articles published in 2011.

**RQ 2. Correlation of the University Academic output between *Google Scholar*, *Scopus* and *Web of Science***

When it comes to compare the academic output of the set of universities, we can observe a strong and positive correlation between *WoS* and *Scopus* over the period (Table 6), both for the first iteration (2014) and for second iteration (2018). However, the productivity data provided by *GS*, despite achieving a positive and significant (α <0.1) correlation coefficient with *WoS* and *Scopus*, achieves weaker values. In addition to this, we can observe lower values for recent years, and lower values in the second iteration.

**Table 6. Correlation (Spearman) of the academic productivity between *Google Scholar*, *Scopus* and *Web of Science* (n=100)**

| Year | *WoS* vs *Scopus* | | *WoS* vs *Google Scholar* | | *Scopus* vs *Google Scholar* | |
|---|---|---|---|---|---|---|
| | 2014 | 2018 | 2014 | 2018 | 2014 | 2018 |
| **2000** | **0.99 | **0.99 | **0.70 | **0.59 | **0.69 | **0.58 |
| **2001** | **0.98 | **0.98 | **0.66 | **0.56 | **0.65 | **0.54 |
| **2002** | **0.98 | **0.97 | **0.65 | **0.56 | **0.54 | **0.57 |
| **2003** | **0.98 | **0.99 | **0.63 | **0.55 | **0.62 | **0.54 |
| **2004** | **0.99 | **0.99 | **0.61 | **0.59 | **0.60 | **0.60 |
| **2005** | **0.99 | **0.99 | **0.65 | **0.63 | **0.65 | **0.61 |
| **2006** | **0.99 | **0.99 | **0.64 | **0.61 | **0.64 | **0.60 |
| **2007** | **0.99 | **0.99 | **0.66 | **0.61 | **0.66 | **0.61 |
| **2008** | **0.99 | **1.00 | **0.64 | **0.57 | **0.64 | **0.58 |
| **2009** | **0.99 | **1.00 | **0.62 | **0.52 | **0.63 | **0.52 |
| **2010** | **0.99 | **0.99 | **0.65 | **0.45 | **0.65 | **0.47 |
| **2011** | **0.99 | **0.98 | **0.56 | **0.47 | **0.56 | **0.50 |
| **2012** | **0.99 | **1.00 | **0.53 | **0.50 | **0.54 | **0.51 |
| **2013** | **0.98 | **0.99 | **0.52 | **0.46 | **0.53 | **0.45 |

The difference between the similarity between controlled and elitist databases (*WoS* and *Scopus*) and dissimilarity with the uncontrolled database (*GS*) can also be observed in the scatter plot offered in Figure 2. Furthermore, we can also check a decrease of the coefficient of determination from the first iteration ($R^2$= 0.46) to the second iteration ($R^2$=0.35) when comparing *WoS* against *GS*.

**Figure 2. Scatter plot of the academic productivity between bibliographic databases**
**(A1) Between *Web of Science* and *Scopus* (first iteration); (A2) Between *Web of Science* and *Scopus* (second iteration); (B1) Between *Web of Science* and *Google Scholar* (first iteration); (B2) Between *Web of Science* and *Google Scholar* (second iteration)**

The misalignment of *GS* with *WoS* and *Scopus* can also be verified at the institutional level. Figure 3 contains the evolution of the productivity of two universities (*Istanbul University* and *Hacettepe University*) from 2000 to 2013. We can observe lower values from *GS* data, especially in the second iteration (negative retroactive growth). While *Istanbul University* seems to fade from GS, *Haceteppe* experiences a significant retroactive growth in 2013, although below the results from *WoS* and *Scopus*.

**Figure 3. Evolution of productivity (from 2000 to 2013) for *Istanbul University* (A) and *Hacettepe University* (B), according to *Google Scholar*, *Scopus* and *Web of Science* in two different iterations (2014 and 2018)**

**RQ 3. University websites as primary versions in *Google Scholar* for publications indexed in *Web of Science***

The coverage of documents published by the fourteen research-focused universities and indexed in *GS* is elevated (84%), although we find great differences among institutions (from 23.2% for *Gebze Institute of Technology* to 95.7% for *Bogaziçi University* (Table 7).

In the particular case of *Istanbul University*, data shows that 85.6% of documents co-authored at least by one author affiliated to this university are indexed in *GS*. However, when it comes to verify whether these documents are already indexed in the university website, we can observe that out of the 1997 documents indexed in GS, only 3 documents are hosted in the official website (all of them as primary version).

This situation does not differ so much in the remaining institutions. As regards the primary versions, the maximum value achieve corresponds to 33 documents while for supplementary versions the maximum is 141 (both for *Izmir Institute of Technology*). And two universities do not provide any document, neither as primary version nor secondary version.

**Table 7. Presence of universities as primary versions in *Google Scholar* for publications indexed in *Web of Science* (2013)**

| UNIVERSITY | WOS | *GOOGLE SCHOLAR* | | | | | | |
| | | INDEX | | | | CITATION | VERSIONS | |
| | Publications | Indexed | % | No Indexed | % | Records | Primary | Secondary |
|---|---|---|---|---|---|---|---|---|
| Istanbul | 2332 | 1997 | 85.6 | 335 | 14.4 | 310 | 3 | 0 |
| Ankara | 1441 | 1243 | 86.3 | 198 | 13.7 | 186 | 12 | 9 |
| Ege | 1548 | 1392 | 89.9 | 156 | 10.1 | 146 | 0 | 13 |
| Middle East | 1458 | 1370 | 94 | 88 | 6 | 60 | 9 | 50 |
| Bogaziçi | 681 | 652 | 95.7 | 29 | 4.3 | 20 | 2 | 45 |
| Istanbul Tech. | 1302 | 1235 | 94.9 | 67 | 5.1 | 33 | 0 | 41 |
| Hacettepe | 2079 | 1775 | 85.4 | 304 | 14.6 | 293 | 6 | 51 |
| Gülhane | 1054 | 923 | 87.6 | 131 | 12.4 | 125 | 0 | 0 |
| Marmara | 1020 | 876 | 85.9 | 144 | 14.1 | 133 | 1 | 16 |
| Gazi | 1611 | 1381 | 85.7 | 230 | 14.3 | 213 | 7 | 2 |
| Erciyes | 1012 | 914 | 90.3 | 98 | 9.7 | 94 | 0 | 0 |
| Gebze | 1317 | 306 | 23.2 | 1011 | 76.8 | 11 | 1 | 7 |
| Izmir | 279 | 261 | 93.5 | 18 | 6.5 | 8 | 33 | 141 |
| Dokuz Eylül | 1040 | 900 | 86.5 | 140 | 13.5 | 127 | 1 | 6 |
| **TOTAL** | **18174** | **15225** | **84** | **2949** | **16** | **1759** | **75** | **381** |

## 6. Discussion and conclusions

**RQ1. Retroactive growth rate**

The retroactive growth at the institution-level appears in all bibliographic databases. In the *WoS* it is of reduced value, growing in the most recent years of the temporal range (2010 to 2013). In *Scopus* the retroactive growth is greater, especially for the first years of the period (2000 to 2003). In the case of *GS*, the retroactive growth is unpredictable, depending on the performance of specific universities. This issue makes the calculation of this parameter hard to evaluate within the environment of a national academic system. In this case, when outlier observations (*Istanbul University*) were removed, the average retroactive growth rate was higher than *WoS* and *Scopus*, as it was expected due to the higher coverage of this database.

These results should be taken however with caution since the productivity of institutions exhibit statistical shortcomings. There are universities with null productivity (In 2000, *WoS* shows 0 publications for 27 universities, whereas *Scopus* 28, and *GS* up to 53). Obviously, with 0 publications measured in the two measurement iterations, the retroactive growth rate is equal to 0; and a change from 0 to 1 represents an increase of 100% for the corresponding institution.

**RQ2. Correlation of Productivity between bibliographic databases**

The number of publications indexed in *GS* and deposited within the official university websites correlates moderately with the number of publications published by the universities and indexed in *WoS* and *Scopus*. Over the years, this correlation ranges from 0.5 to 0.7 in the case of *Scopus*, and from 0.5 to 0.7 in the case of *WoS* in the first measurement iteration (2014), decreasing slightly in the second iteration (up to 0.6).

Orduna-Malea, Serrano-Cobos and Lloret-Romero (2009) obtained a slightly higher correlation (R= 0.7) between *GS* and *Scopus* for the Spanish academic system (using Pearson and removing outliers as well). However, results cannot be compared directly due to the differences (different years of measurement and different academic context).

The disparity of the productivity data provided by *GS* can be due to different factors:

a) Technical reasons: due to indexing errors in *GS* (lack of coverage), errors search filter operating (documents without year of publication that are non-recoverable) and lack of exhaustiveness of the 'site' command.
b) Conceptual reasons: due to the comparison of non-comparable documents. Under the academic web domain it can appear not only a wide variety of document typologies (not only articles published in journals), but also material not necessarily written by university authors.
c) Reasons related to the commitment with Open Access practices: due to the adoption of self-archiving practices, use of the institutional repository and open science national or institutional policies.

**RQ3. Coverage of publications in official university websites**

In the particular case of Turkey, this work evidences an invisibility of publications indexed in *WoS* and co-authored by authors affiliated to the analyzed universities, in the official academic websites. This fact negatively affects the visibility of universities on the Web in general, and on *GS* in particular.

A large percentage of the documents published by the universities have been located in *GS* but in other sources different from those of the universities (See Table 7). This may be due to the fact that researchers put and share their work on platforms such as thematic repositories (for example, *Arxiv*), academic social networks (for example, *ResearchGate*) or personal websites. Therefore, the reason for the detected invisibility may be due to a factor of a lack of commitment to Open Access in the particular academic environments studied.

**Final Remarks**

This work proves that the url-based method to calculate institutional productivity in *GS* is not a good proxy for the total number of publications indexed in *WoS* and *Scopus*. However, the main reason is not directly related to the operation of *GS* but with a lack of commitment of universities with open access and the use of institutional repositories.

The practices of the authors, shaped by the editorial, university and governmental policies, have a direct reflection on the results obtained in this work. Its replication in other environments with other policies and academic practices could serve to verify and reinforce these conclusions. Similarly, the development of Plan S (https://www.coalition-s.org) could change the situation in many national university environments. For this reason, even when the url-based method of *GS* is not the most accurate, it could constitute a barometer of the evolution of university policies around open access.

## 7. References

Aguillo, I. F. (2012). Is Google Scholar useful for bibliometrics? A webometric analysis. *Scientometrics, 91*(2), 343-351. https://doi.org/10.1007/s11192-011-0582-8

Aguillo, I. F., Ortega, J. L., & Fernández, M. (2008). Webometric ranking of world universities: Introduction, methodology, and future developments. *Higher education in Europe, 33*(2-3), 233-244. https://doi.org/10.1080/03797720802254031

Amara, N., Landry, R., & Halilem, N. (2015). What can university administrators do to increase the publication and citation scores of their faculty members? *Scientometrics, 103*(2), 489-530. https://doi.org/10.1007/s11192-015-1537-2

Arlitsch, K., & O'Brian, P. S. (2012). Invisible institutional repositories: Addressing the low indexing ratios of IRs in Google. *Library Hi Tech, 30*(1), 60–81. https://doi.org/10.1108/07378831211213210

Aytac, S. (2010). *An examination of international scientific collaboration in a developing country (Turkey) in the post Internet era*. Brookville, NY: Long Island University.

De Winter, J. C., Zadpoor, A. A., & Dodou, D. (2014). The expansion of Google Scholar versus Web of Science: a longitudinal study. *Scientometrics, 98*(2), 1547-1565. https://doi.org/10.1007/s11192-013-1089-2

Delgado López-Cózar, E.; Orduna-Malea, E.; Martín-Martín, A. (2019). Google Scholar as a Data Source for Research Assessment. In W. Glänzel, H.F. Moed, U. Schmoch, & M. Thelwall (Eds.). *Springer handbook of science and technology indicators*. Heidelberg: Springer.

Franceschini, F., Maisano, D., & Mastrogiacomo, L. (2016a). Empirical analysis and classification of database errors in Scopus and Web of Science. *Journal of Informetrics, 10*(4), 933-953. https://doi.org/10.1016/j.joi.2016.07.003

Franceschini, F., Maisano, D., & Mastrogiacomo, L. (2016b). The museum of errors/horrors in Scopus. *Journal of Informetrics, 10*(1), 174-182. https://doi.org/10.1016/j.joi.2015.11.006

Gusenbauer, M. (2019). Google Scholar to overshadow them all? Comparing the sizes of 12 academic search engines and bibliographic databases. *Scientometrics, 118*(1), 177–214. https://doi.org/10.1007/s11192-018-2958-5

Harzing, A. W. (2013). A preliminary test of Google Scholar as a source for citation data: a longitudinal study of Nobel prize winners. *Scientometrics, 94*(3), 1057-1075. https://doi.org/10.1007/s11192-012-0777-7

Harzing, A. W. (2014). A longitudinal study of Google Scholar coverage between 2012 and 2013. *Scientometrics, 98*(1), 565–575. https://doi.org/10.1007/s11192-013-0975-y

Hook, D. W., Porter, S. J., & Herzog, C. (2018). Dimensions: Building context for search and evaluation. *Frontiers in Research Metrics and Analytics*, 3, 23. https://doi.org/10.3389/frma.2018.00023

Jacsó, P. (2010). Metadata mega mess in Google Scholar. *Online Information Review, 34*(1), 175-191. https://doi.org/10.1108/14684521011024191

Martín-Martín, A., Orduna-Malea, E., Ayllón, J. M., & Delgado López-Cozar, E. (2016). A two-sided academic landscape: snapshot of highly-cited documents in Google Scholar (1950-2013). *Revista española de documentación científica, 39*(4), 1-21. http://dx.doi.org/10.3989/redc.2016.4.1405

Mingers, J., & Meyer, M. (2017). Normalizing Google Scholar data for use in research evaluation. *Scientometrics, 112*(2), 1111-1121. https://doi.org/10.1007/s11192-017-2415-x

Mingers, J., O'Hanley, J. R., & Okunola, M. (2017). Using Google Scholar institutional level data to evaluate the quality of university research. *Scientometrics, 113*(3), 1627-1643. https://doi.org/10.1007/s11192-017-2532-6

Mongeon, P., & Paul-Hus, A. (2016). The journal coverage of Web of Science and Scopus: a comparative analysis. *Scientometrics, 106*(1), 213-228. https://doi.org/10.1007/s11192-015-1765-5

Moskovkin, V. M. (2009). The potential of using the Google Scholar search engine for estimating the publication activities of universities. *Scientific and technical information processing, 36*(4), 198-202. https://doi.org/10.3103/S0147688209040029

Moskovkin, V. M., Delux, T., & Moskovkina, M. V. (2012). Comparative Analysis of University Publication Activity by Google Scholar:(On Example of Leading Czech and Germany Universities). *Cybermetrics: International Journal of Scientometrics, Informetrics and Bibliometrics, 16*(1), 1-9. http://hdl.handle.net/10261/174558

Orduna-Malea, E., & Delgado López-Cózar, E. (2015). The dark side of Open Access in Google and Google Scholar: the case of Latin-American repositories. *Scientometrics, 102*(1), 829-846. https://doi.org/10.1007/s11192-014-1369-5

Orduna-Malea, E., & Delgado-López-Cózar, E. (2018). Dimensions: re-discovering the ecosystem of scientific information. *El Profesional de la Información, 27*(2), 420-431. https://doi.org/10.3145/epi.2018.mar.21

Orduna-Malea, E., Ayllon, J. M., Martín-Martín, A., & Delgado López-Cózar, E. (2017). The lost academic home: Institutional affiliation links in Google Scholar Citations. *Online Information Review, 41*(6), 762-781. https://doi.org/10.1108/OIR-10-2016-0302

Orduña-Malea, E., Martín-Martín, A., Ayllón, Juan M. & Delgado López-Cózar, E. (2016). *La revolución Google Scholar: Destapando la caja de Pandora académica*. Granada: UNE.

Orduna-Malea, E., Martín-Martín, A., Delgado López-Cozar, E. (2017). Google Scholar as a source for scholarly evaluation: a bibliographic review of database errors. *Revista española de documentación científica, 40*(4), 1-33. https://doi.org/10.3989/redc.2017.4.1500

Orduña-Malea, E., Serrano-Cobos, J. & Lloret-Romero, N. (2009). Las Universidades públicas españolas en Google Scholar: presencia y evolución en su publicación académica web. *El profesional de la información, 5*(18), 493-501. https://doi.org/10.3145/epi.2009.sep.02

Ortega, J. L. (2014). *Academic search engines: A quantitative outlook*. Oxford: Chandos Publishing.

Ramsden, P. (1994). Describing and explaining research productivity. *Higher Education, 28*(2), 207-226. https://doi.org/10.1007/BF01383729

Ranjbar-Sahraei, B., van Eck, N. J., & de Jong, R. (2018). Accuracy of affiliation information in Microsoft Academic: Implications for institutional level research evaluation. In R. Costas, T. Franssen, & A. Yegros-Yegros (Eds.), *STI 2018 Conference Proceedings: Proceedings of the 23rd International Conference on Science and Technology Indicators* (pp. 1065-1067). Leiden, The Netherlands: Centre for Science and Technology Studies (CWTS), Leiden University. http://hdl.handle.net/1887/65339

**Figure 1. Overview of the general limitations of Title Search in *Google Scholar***

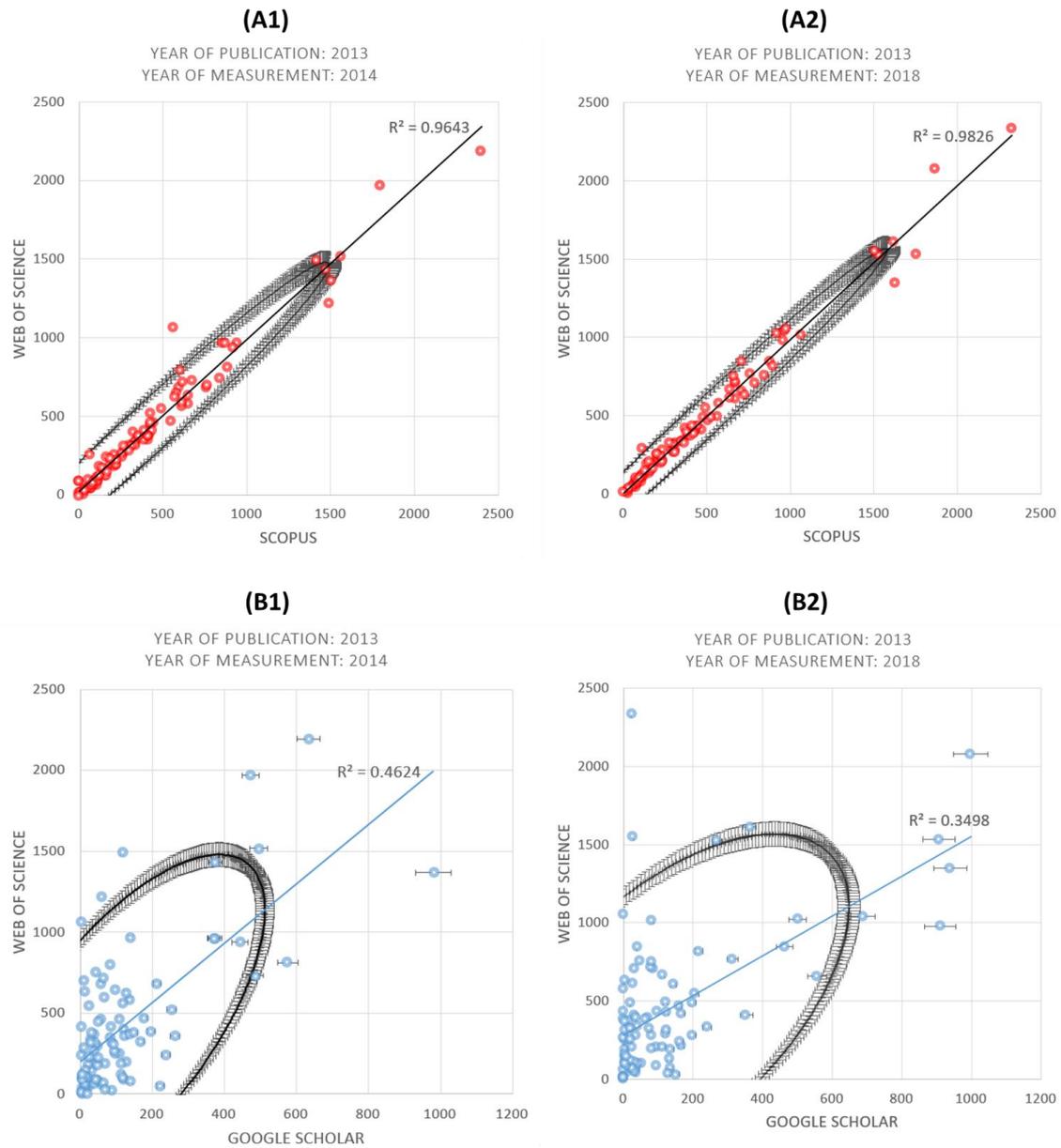

**Figure 2. Scatter plot of the academic productivity between bibliographic databases**
**(A1)** Between *Web of Science* and *Scopus* (first iteration); **(A2)** Between *Web of Science* and *Scopus* (second iteration); **(B1)** Between *Web of Science* and *Google Scholar* (first iteration); **(B2)** Between *Web of Science* and *Google Scholar* (second iteration)

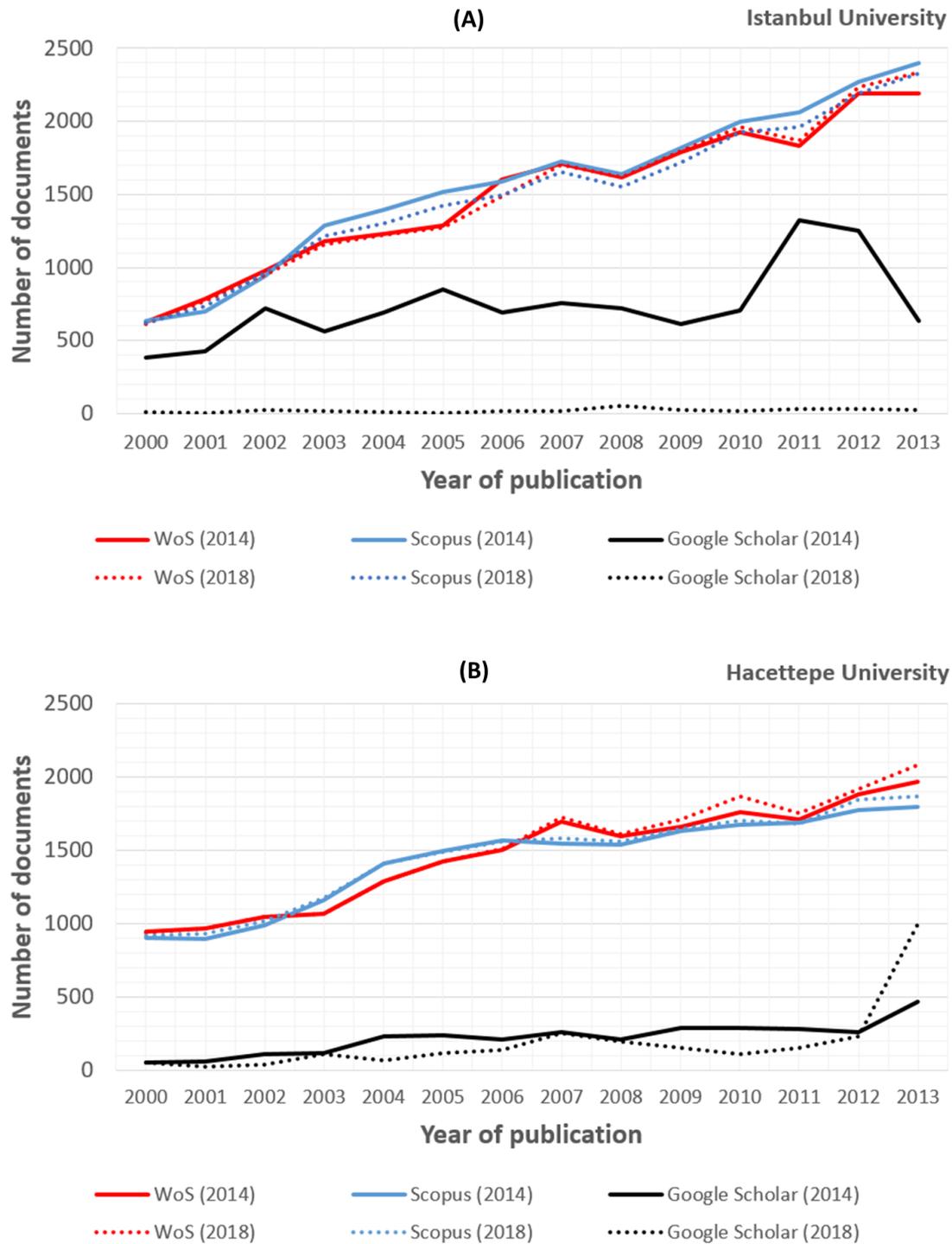

**Figure 3. Evolution of productivity (from 2000 to 2013) for *Istanbul University* (A) and *Hacettepe University* (B), according to *Google Scholar*, *Scopus* and *Web of Science* in two different iterations (2014 and 2018)**